\documentclass[10pt,letterpaper]{article}
\usepackage{M11}
\usepackage{slashed,bbm}

\setlist{leftmargin=5mm} %

\numberwithin{equation}{section}

\renewcommand{\tilde}[1]{\mkern 2mu\widetilde{\mkern-1mu#1\mkern-1mu}\mkern 2mu}

\providecommand{\tabularnewline}{\\}

\usepackage{graphicx}
\usepackage{feynmp}
\usepackage{color}
\DeclareGraphicsExtensions{.pdf}  %

\def \bea  {\begin{eqnarray}}
\def \eea  {\end{eqnarray}}

\newcommand{\bra}[1]{\langle{#1}|}
\newcommand{\ket}[1]{|{#1}\rangle}

\newcommand{\nn}{\nonumber}

\newcommand{\ns} \normalsize

\newcommand{\5}{$AdS_5\times S^5$}

\newcommand{\2}{$AdS_2\times S^2\times T^6$}
\renewcommand{\ads}{\textrm{AdS}^5\times \textrm{S}^5}

\AtBeginDocument{
  
}

\makeatother

\begin{document}

\setpreprint{MIFPA-13-25, QGaSLAB-13-07}

\title{
Scattering in AdS$_\mathsf{2}$/CFT$_\mathsf{1}$ and the BES Phase
}

\author{Michael C. Abbott,\alabel{1} Jeff Murugan,\alabel{\, 1} Per Sundin,\alabel{1} 
and Linus Wulff\alabel{\; 2}
\address[1]{The Laboratory for Quantum Gravity \& Strings,\\ 
Department of Mathematics \& Applied Mathematics,\\
University of Cape Town, Rondebosch 7701, South Africa }
\address[2]{George P.\ \& Cynthia Woods Mitchell Institute for Fundamental Physics and Astronomy, \\
Texas A\&M University, College Station, TX 77843, USA}
\address{michael.abbott@uct.ac.za,
jeff@nassp.uct.ac.za, nidnus.rep@gmail.com, linus@physics.tamu.edu
}}

\date{6 August 2013}  %

\maketitle

\begin{abstract}
We study worldsheet scattering for the type IIA superstring in $AdS_2\times S^2 \times T^6$. 
Using the Green--Schwarz action to quartic order in fermions we take the near-BMN limit,
where as in the AdS$_3$/CFT$_2$ case 
there are both massive and massless excitations. For the massive excitations we compute all possible tree-level processes, and show that these 
agree with a truncated version of the exact $AdS_5\times S^5$ S-matrix. 
We also compute several S-matrix elements 
involving massless excitations. 
At one loop we find that the dressing phase is the same Hern\'{a}ndez--L\'{o}pez phase appearing in AdS$_5$/CFT$_4$. 
We see the same phase when calculating this by semiclassical means using the $PSU(1,1|2)/U(1)^2$ coset sigma model,
for which we can also study the scattering of fermions. 
This supports the conjecture that the all-loop dressing phase is again the BES phase, rather than 
a new phase like that seen in AdS$_3$/CFT$_2$. 
\end{abstract}

\section{Introduction}

The original AdS/CFT correspondence relates type IIB strings in $AdS_{5}\times S^{5}$
to $\mathcal{N}=4$ super-Yang--Mills theory, both of which are maximally
supersymmetric with 32 supercharges \cite{Maldacena:1997re}. Our best understanding of
states not protected by supersymmetry comes from the fact that both
theories are integrable, which allowed a description valid at
all values of the 't Hooft coupling $\lambda$ \cite{Minahan:2002ve,Bena:2003wd,Arutyunov:2009ga,Beisert:2010jr}.
One of the central objects in this picture is the S-matrix \cite{Beisert:2005tm}
and the dressing phase \cite{Arutyunov:2004vx,Hernandez:2006tk,Beisert:2006ib,Beisert:2006ez,Beisert:2007hz},
a scalar factor needed to match the string and gauge theory (spin-chain) equations. While one is generally able to determine the S-matrix via the underlying symmetries of the problem, the dressing phase is more complicated. 
Nevertheless, it is known to all-loop under the acronym BES or BES/BHL phase \cite{Beisert:2006ib,Beisert:2006ez}.
The one-loop term was re-constructed by semiclassical means in \cite{Gromov:2007cd,Chen:2007vs},
and the phase was observed directly in worldsheet scattering by \cite{Klose:2006zd,Klose:2007rz} up
to two loops.

The first extension of these ideas to theories with less supersymmetry
was the case of IIA strings in $\smash{AdS_{4}\times CP^3}$ and the dual ABJM gauge theory
\cite{Aharony:2008ug,Minahan:2008hf,Gromov:2008qe,Klose:2010ki},
with 24 supercharges. While this new version of the correspondence has a variety of new features, 
the dressing phase turned out to be nothing but half the BES phase \cite{Gromov:2008qe},
leading to the idea that the phase might be universal. This phase has now been 
confirmed at weak coupling by an eight-loop calculation \cite{Mauri:2013vd}. 

More recently there has been much interest in $AdS_{3}\times S^{3}\times T^{4}$
and the related $AdS_{3}\times S^{3}\times S^{3}\times S^{1}$ background
\cite{David:2008yk,Babichenko:2009dk,Zarembo:2010yz,David:2010yg,OhlssonSax:2011ms,Forini:2012bb,Rughoonauth:2012qd,Sundin:2012gc,Cagnazzo:2012se,Sax:2012jv,Ahn:2012hw,Borsato:2012ud,Abbott:2012dd,Beccaria:2012kb,Borsato:2012ss,Beccaria:2012pm,Sundin:2013ypa,Borsato:2013qpa,Borsato:2013hoa,Abbott:2013mpa,Sundin:2013uca,Hoare:2013pma,Hoare:2013ida,Hoare:2011fj,Bianchi:2013nra,Engelund:2013fja},
with 16 supersymmetries. The gauge theories dual to these  backgrounds are not
well understood, but despite this some all-loop progress has been
made using integrability.  Here the description in terms of a supercoset sigma model only 
captures the curved directions, omitting 4 (or 1) flat directions, and this 
limitation is inherited by the integrable description. However, at least at the classical level, the integrability seems to extend to the full superstring \cite{Sundin:2012gc} (see also \cite{Sax:2012jv} another attempt at incorporating the flat directions).  
It has emerged that the dressing phase is not the standard BES phase, starting at one loop, 
as seen in direct
semiclassical calculations \cite{David:2010yg,Beccaria:2012kb,Abbott:2013mpa},
in worldsheet scattering calculations \cite{Rughoonauth:2012qd,Sundin:2013ypa}, various 
indirect ways \cite{Abbott:2012dd,Beccaria:2012pm}, and in the all-loop proposal of  
\cite{Borsato:2013hoa}. 

In this paper we study strings on an even less supersymmetric background
with only 8 supercharges, namely $AdS_{2}\times S^{2}\times T^{6}$
\cite{Berkovits:1999zq,Zhou:1999sm,Sorokin:2011rr,Cagnazzo:2011at,Cagnazzo:2012uq,Murugan:2012mf}.
This gives an embedding of the $AdS_{2}\times S^{2}$ near-horizon geometry
of the 4-dimensional extremal Reissner--Nordström black hole into critical string
theory, and it is thought to be dual to a superconformal quantum mechanics (CFT$_1$) \cite{Claus:1998ts,BrittoPacumio:1999ax,Chamon:2011xk}. We will approach this from the point of view of integrability \cite{Sorokin:2011rr,Cagnazzo:2011at}, with one of our main goals being to show that the dressing phase is like that of $AdS_{5}$. This was conjectured in \cite{Sorokin:2011rr}
but has so far not been checked beyond tree level.  Note that in this case the supercoset sigma model represents only a (classically consistent) truncation of the full superstring \cite{Sorokin:2011rr} and many of the standard integrability tools are therefore not applicable without some modification.

We begin by setting up the Green--Schwarz action, extending \cite{Murugan:2012mf}
to fourth order in the fermions using \cite{Wulff:2013kga}. There
are in fact several $AdS_{2}\times S^{2}\times T^{6}$ backgrounds
related by T-duality \cite{Sorokin:2011rr}, and we focus on the
type IIA example supported by non-zero RR flux $F_{2}$ through $AdS_{2}$ (and $F_{4}$
flux through $S^{2}\times T^{6}$), as in \cite{Murugan:2012mf}. We then take the near BMN limit, expanding
the action up to to sixth order (but keeping only fermions up to quartic order) which is sufficient
for one-loop scattering calculations when the external legs are bosonic. As in $AdS_{3}$ there is a division 
between the massive modes, which can be described by a coset model, and the massless modes from the flat torus directions  (together with corresponding fermions).

Our results are as follows: \vspace{-2mm}
\begin{itemize}
\item At the tree-level we compute all possible scattering amplitudes with
 massive asymptotic states. These results agree with a truncation of
the exact \5 S-matrix
of \cite{Beisert:2005tm}, as is natural since the classical action
can be consistently truncated to $AdS_{2}\times S^{2}$. We also consider
the case where we have asymptotic states mixing massive and massless
particles. While most such amplitudes are zero, we see that there
are certain states that give non-zero results.
\item At one loop we see results consistent with the original Hern\'{a}ndez--L\'{o}pez 
phase\footnote{There are some technical issues for the one-loop diagrams. Due to a known breakdown of the light-cone gauge fixing at one loop, some of the amplitudes fail to be UV finite. While the one-loop phase still can be extracted unambiguously, we nevertheless compute the remaining terms in the near-flat-space limit \cite{Sundin:2013ypa,Engelund:2013fja}.} \cite{Hernandez:2006tk}. We also observe that truncating the model to just the massive modes  breaks the agreement with the HL phase.
\item We also compute the phase semiclassically, using the coset model and the algebraic curve, following \cite{Chen:2007vs,David:2010yg,Abbott:2013mpa}.
This calculation contains only the massive modes, but nevertheless
agrees with the result from the full worldsheet theory, not the truncation to
the massive sector.
\item As a cross-check we also find certain terms of the one-loop amplitudes
in terms of the tree-level results, using the optical theorem and
generalised unitarity. For these terms we see perfect agreement. 
\end{itemize}

\subsubsection{Outline}

Section \ref{stringsection} sets up the action and the BMN expansion. In sections \ref{sec:tree-level} and \ref{sec:one-loop} we study worldsheet scattering amplitudes at tree level and one loop, and connections between these. We then turn  to a semiclassical derivation of the one-loop phase in section \ref{sec:semiclass-HL} and end with a summary and discussion in section  \ref{sec:Conclusions}. In the appendix we %
demonstrate that the massive tree-level S-matrix agrees with a truncation of the $AdS_3\times S^3\times T^4$ S-matrix.

\section{The Green--Schwarz Superstring in $AdS_2\times S^2\times T^6$}
\label{stringsection}
The type IIA and IIB Green--Schwarz (GS) string is known in a closed form in a general supergravity background up to quartic order in fermions \cite{Wulff:2013kga}. We will consider the type IIA \2 solution here since we can combine the two Majorana--Weyl spinors of the type IIA superspace into a single 32-component Majorana spinor, simplifying somewhat our analysis. The type IIB \2 solutions and type IIA solutions with different RR-flux are related to each other by T-dualities along the toroidal directions, see \cite{Sorokin:2011rr} for a detailed discussion. 

The type II Green--Schwarz superstring action in a general supergravity background can be expanded in the number of fermions as follows:
\begin{equation}
\label{action}
S=\hat g\int d^2\sigma\,\mathcal L\,,\qquad \mathcal L=\mathcal L^{(0)}+\mathcal L^{(2)}+\mathcal L^{(4)}+\ldots\,.
\end{equation}
Since we anticipate a dual CFT we write the coupling as $\hat g = \sqrt{\lambda}/4\pi$. Since there is no NS-NS flux the purely bosonic Lagrangian is simply
\begin{equation}
\label{eq:L0}
\mathcal L^{(0)}=\frac{1}{2}\gamma^{ij}e_i{}^Ae_j{}^B\eta_{AB}\,,\qquad\gamma^{ij}=\sqrt{-\det g\,} g^{ij}\,,
\end{equation}
where $e_i{}^A(X)$ $(A=0,1,\cdots,9)$ are the vielbeins of the bosonic part of the background pulled back to the worldsheet and $g_{ij}$ is an independent worldsheet metric. We write the metric of \2 in terms of global coordinates like \cite{Rughoonauth:2012qd}
\begin{equation}
ds^2=-\Big(\frac{1+\frac{1}{4}x_1^2}{1-\frac{1}{4}x_1^2}\Big)^2dt^2+\frac{dx_1^2}{\big({1-\frac{1}{4}x_1^2}\big)^2}+
\Big(\frac{1-\frac{1}{4}x_2^2}{1+\frac{1}{4}x_2^2}\Big)^2d\varphi^2+\frac{dx_2^2}{\big({1+\frac{1}{4}x_2^2}\big)^2}+dx_{a'}^2\,,
\end{equation}
where $a'=3,\dots,8$ denotes the $T^6$ directions. The spin connection of the background is readily calculated from the vanishing of the torsion,
\[
de^A+e^B \omega_B{}^A=0
\]
which gives
\begin{equation}
\label{eq:spin-connection-rep}
\omega^{01}=-\frac{x_1 dt}{1-\frac{1}{4}x_1^2},\qquad \omega^{23}=-\frac{x_2 d\varphi}{1+\frac{1}{4}x_2^2}\,.
\end{equation}

The terms quadratic in fermions take the form \cite{Cvetic:1999zs}
\begin{equation}
\label{eq:L2}
\mathcal L^{(2)}=ie_i{}^A\,\bar\Theta\Gamma_AK^{ij}{\mathcal D}_j\Theta\,,\qquad K^{ij}=\gamma^{ij}-\varepsilon^{ij}\Gamma_{11}\,.
\end{equation}
The appearance of the matrix $K^{ij}$ is related to kappa symmetry and the Killing spinor derivative $\mathcal D$ is given below. Furthermore, the quartic fermion terms in the action take the form \cite{Wulff:2013kga}\footnote{Our normalization of $\Theta$ differs from that of \cite{Wulff:2013kga} by a factor of $\sqrt2$.}
\begin{align}
\label{eq:L4}
\mathcal L^{(4)} =&
-\frac{1}{2}\bar\Theta\Gamma^A\mathcal{D}_i\Theta\,\bar\Theta\Gamma_A K^{ij}\mathcal{D}_j\Theta
+\frac{i}{6}e_i{}^A\,\bar\Theta\Gamma_AK^{ij}\mathcal M\mathcal D_j\Theta
+\frac{i}{48}e_i{}^Ae_j{}^B\,\bar\Theta\Gamma_AK^{ij}\big(M+\tilde M\big)S\Gamma_B\Theta
\nonumber\\
&{}
+\frac{1}{48}e_i{}^Ae_j{}^B\,\bar\Theta\Gamma_A{}^{CD}K^{ij}\Theta\,\big(3\bar\Theta\Gamma_BU_{CD}\Theta-2\bar\Theta\Gamma_CU_{DB}\Theta\big)
\nonumber\\
&{}
-\frac{1}{48}e_i{}^Ae_j{}^B\,\bar\Theta\Gamma_A{}^{CD}\Gamma_{11}K^{ij}\Theta\,\big(3\bar\Theta\Gamma_B\Gamma_{11}U_{CD}\Theta+2\bar\Theta\Gamma_C\Gamma_{11}U_{DB}\Theta\big)\,.
\end{align}
where the definition of $\mathcal D$, $\mathcal M$, $M$ and $U_{AB}$ for a general type II supergravity background can be found in \cite{Wulff:2013kga}. Here we will focus on the \2 type IIA solution with non-zero $F_2$ and $F_4$ flux,
\[\begin{aligned}
 F_2 &=\frac{e^{-\phi}}{2}e^b e^a \varepsilon_{ab}, \\ 
 F_4 &=-\frac{e^{-\phi}}{2}e^{\hat b} e^{\hat a} \varepsilon_{\hat a\hat b} J\,,
\end{aligned}\]
where $a,b=0,1$, $\hat a,\hat b=2,3$ refer to $AdS_2$ and $S^2$ respectively, $\varepsilon^{01}=1=\varepsilon^{23}$ and $J$ is the K\"ahler form on $T^6$ which we take to be
\begin{equation}
J=-dx^4dx^5-dx^6dx^7-dx^8dx^9\,.
\end{equation}
Note that this choice of K\"ahler form breaks the original $SO(6)$ invariance of the $T^6$ down to $U(3)$. There are also other possible combinations of RR-flux for the \2 string and the different solutions are related by T-duality \cite{Sorokin:2011rr}. 

Using the above the combined flux can be written as
\[
\label{eq:F-slash}
S=e^\phi\left(\frac12F_{AB}\Gamma^{AB}\Gamma_{11}+\frac{1}{4!}F_{ABCD}\Gamma^{ABCD}\right)
=-4\mathcal{P}_8\Gamma^{01}\Gamma_{11}\,,
\]
where $\mathcal{P}_8$ is a projector that projects on the eight supersymmetries preserved by the background. Explicitly it is given by
\bea
\label{eq:P-projector}
\mathcal{P}_8=\frac{1}{8}\big(2-i\slashed J\gamma_7\big)\,,\qquad \slashed J=J_{a'b'}\Gamma^{a'b'}=2\big(\Gamma^4\Gamma^5+\Gamma^6\Gamma^7+\Gamma^8\Gamma^9\big)\,,\qquad\gamma_7=i\Gamma^4\cdots\Gamma^9
\eea
and it commutes with the gamma-matrices of the $AdS_2\times S^2$ part of the background, $[\mathcal{P}_8,\Gamma_a]=[\mathcal{P}_8,\Gamma_{\hat a}]=0$.

Furthermore, $\Theta$ is taken to be a 32-component Majorana spinor and the Killing spinor derivative is given by
\begin{equation}
\label{E}
{\mathcal D_i}\Theta=(\partial_i-\frac{1}{4}\omega_i{}^{AB}\Gamma_{AB}+\frac{1}{8}e_i{}^A\,S\Gamma_A)\Theta\,.
\end{equation}
The remaining objects appearing in (\ref{eq:L4}) reduce, in \2, to
\bea
\label{eq:quartic-matrices}
&&{}U_{AB}=\frac{1}{32}S\Gamma_{[A}S\Gamma_{B]}-\frac{1}{4}R_{AB}{}^{CD}\Gamma_{CD}\,,
\nonumber\\
&&{}M^\alpha{}_\beta=
\frac{i}{16}\bar\Theta S\Theta\,\delta^\alpha_\beta
-\frac{i}{8}\Theta^\alpha\,(\bar\Theta S)_\beta
+\frac{i}{8}(\Gamma^AS\Theta)^\alpha\,(\bar\Theta\Gamma_A)_\beta\,,\qquad \Tilde M=\Gamma_{11}M\Gamma_{11}\,,
\nonumber\\
&&{}\mathcal M^\alpha{}_\beta=M^\alpha{}_\beta +\tilde M^\alpha{}_\beta
+\frac{i}{8}(S\Gamma^A\Theta)^\alpha\,(\bar\Theta\Gamma_A)_\beta
-\frac{i}{16}(\Gamma^{AB}\Theta)^\alpha\,(\bar\Theta\Gamma_AS\Gamma_B)_\beta\,,
\eea
where the nonzero components of the Riemann tensor are
\begin{equation}
\label{eq:Riemann}
R_{ab}{}^{cd}=2\delta_{[a}^c\delta^d_{b]}\,,\qquad R_{\hat a\hat b}{}^{\hat c\hat d}=-2\delta_{[\hat a}^{\hat c}\delta^{\hat d}_{\hat b]}\,.
\end{equation}
With this we have all the necessary ingredients for the upcoming analysis. Of course, the action is invariant under bosonic reparameterizations and fermionic kappa symmetry which need to be fixed in order to extract the physical consequences of the theory.

\subsection{The BMN expansion}
We now look at the near-BMN expansion, for a string moving along the $\varphi$-direction of $S^2$ close to the speed of light. Similar analysis for more symmetric versions of $AdS / CFT$ have been performed in \cite{Callan:2004ev,Abbott:2011xp,Rughoonauth:2012qd}.

In order to remove the unphysical fermionic degrees of freedom we will use a standard light-cone kappa symmetry gauge-fixing adapted to the BMN geodesic
\bea
\label{eq:gauges}
\Gamma^+\Theta=0,\qquad \Gamma^\pm=\Gamma^0\pm\Gamma^3\,.
\eea
Furthermore, introducing light-cone coordinates as $x^\pm=\frac{1}{2}(t\pm\varphi)$ the action (\ref{action}) becomes, to quadratic order in coordinates
\begin{align}
\label{eq:L2ferm}
\mathcal{L}_2 &= \frac{1}{2}\gamma^{ij}\partial_i x^M \partial_j x_M - \frac{1}{2}\gamma^{ij} \partial_i x^+\partial_j x^+ (x_1^2+x_2^2\big)
\nn \\
&{} \ \ \
-2i\,\bar\Theta\big(\gamma^{ij}+\varepsilon^{ij}\Gamma_{11}\big)\partial_i x^+\Gamma^-\partial_j\Theta
-2i\gamma^{ij}\partial_ix^+\partial_jx^+\,\bar{\Theta}\Gamma^-\Gamma^1\Gamma_{11}\mathcal P_8\Theta\,.
\end{align}
It is clear from this expression that only the eight supercoset fermions (four after kappa symmetry gauge-fixing), which satisfy $\Theta=\mathcal P_8\Theta$, get a non-zero mass while the other fermions, $\big(1-\mathcal P_8 \big)\Theta$, remain massless.

The bosonic worldsheet parameterization invariance is fixed by \cite{Arutyunov:2005hd, Frolov:2006cc}
\bea
\label{eq:gauge}
x^+=\tau,\qquad p_-=\frac{\delta \mathcal{L}}{\delta \dot{x}^-}=1\,,
\eea
which corresponds to the standard $a=\frac{1}{2}$ light-cone gauge, see \cite{Arutyunov:2009ga} for details.

To lowest order we fix $\gamma^{ij}=\eta^{ij}$ and the quadratic Lagrangian becomes (the worldsheet metric has signature $(+-)$ in our conventions)
\begin{align}
\mathcal{L}_2=
\frac{1}{2}\big(\partial_+x^m\partial_-x_m -x_1^2-x_2^2\big)
-2i\,\bar\Theta_+\Gamma^-\partial_-\Theta_+
-2i\,\bar\Theta_-\Gamma^-\partial_+\Theta_-
+4i\,\bar{\Theta}_+\Gamma^-\Gamma^1\mathcal P_8\Theta_-\,,
\nn 
\end{align}
where $\partial_\pm=\partial_0\pm\partial_1,  \Theta_\pm=\frac{1}{2}(1\pm\Gamma_{11})\Theta$ and $m$ is an index running over the eight, massive and massless, transverse directions. Using a real representation for the $AdS_2\times S^2$ bosons and a complex notation for the remaining $T^6$ coordinates, the quadratic Lagrangian can be written 
\begin{align}
\label{eq:L2-gf}
 \mathcal{L}_2=
 \frac{1}{2}\big(\partial_+ x_r \partial_- x_r+\partial_+ y_i \partial_- \bar y_i+\partial_- y_i \partial_+ \bar y_i-x_r^2\big)+i\,\bar \chi^j_+\partial_- \chi^j_++i\, \bar \chi^j_- \partial_+ \chi^j_- -\bar\chi^1_+ \chi_-^1-\bar \chi^1_- \chi_+ ^1,
\end{align}
where $r=1,2, i=2,3,4$ and $j=1,2,3,4$. (See \cite{Murugan:2012mf} for details.) The complex fields are all charged under the three $U(1)$ symmetries. This is summarised in the following table:
\[ \label{tab:charges}
\begin{tabular}{c|cc|ccc|ccc}
 & \multicolumn{2}{c|}{Coset} &  &  & \multicolumn{2}{c}{Torus} &  & \tabularnewline
 & $x_{1},x_{2}$ & $\chi_{\pm}^{1}$  & $y_{2}$  & $y_{3}$ & $y_{4}$ & $\chi_{\pm}^{2}$  & $\chi_{\pm}^{3}$ & $\chi_{\pm}^{4}$ \tabularnewline
\hline
mass & 1 & 1 & 0 & 0 & 0 & 0 & 0 & 0\tabularnewline
\hline \nn
$U(1)_{2}$  & 0 & $-1/2$ & $-1$ &   &  & $-1/2$ & 1/2  & 1/2 \tabularnewline
$U(1)_{3}$  & 0 & $-1/2$ &  & $-1$ &  & 1/2  & $-1/2$ & 1/2 \tabularnewline
$U(1)_{4}$  & 0 & $-1/2$ &  &  & $-1$ & 1/2  & 1/2  & $-1/2$\tabularnewline
\hline
\end{tabular}
\]

Beyond the quadratic approximation the BMN expansion gives a series in inverse powers of the coupling
\begin{equation}
\mathcal{L}=\mathcal{L}_2+\frac{1}{\hat g}\mathcal{L}_4+ \frac{1}{\hat g^2}\mathcal{L}_6 + \bigodiv{\hat{g}^3}\,.
\end{equation}
Consistency of the gauge-fixing (\ref{eq:gauge}) at higher order in perturbation theory demands that we add sub-leading corrections to the worldsheet metric. The precise form of the corrections is found by looking at the equations of motion for $x^-$ and their leading-order form is\footnote{Note that these corrections are significantly more complicated in the generalized light-cone gauge where $a\neq\frac{1}{2}$.}
\bea
\gamma_{ij}=\eta_{ij}+\frac{1}{\hat g}\hat\gamma_{ij},\qquad \hat\gamma_{00}=\hat\gamma_{11}=-\frac{1}{2}\big(x_1^2-x_2^2\big),\qquad \hat\gamma_{01}=0\,.
\eea
For the sixth order Lagrangian we need to add further $\mathcal{O}(g^{-2})$ corrections. While easily determined via the equations of motions, they nevertheless have a fairly complicated form and we will not present them here. However, and more importantly, the light-cone gauge seems to break down at the one-loop level, at least naively. This is a known problem that has been observed earlier for both the \5 and $AdS_3\times S^3\times T^4$ string in \cite{Sundin:2013ypa}. The breakdown of the gauge fixing results in UV divergent amplitudes for diagonal scattering where all particles carry the same flavor index. We will present a more detailed discussion on this later in the paper. 

Expanding the Lagrangian in (\ref{action}) to quartic order in transverse fields we get
\begin{align}
\label{eq:quarticL}
\hat g \,\mathcal{L}_4  = \ &
\frac{1}{4}\big(\partial_+x_1\partial_- x_1\,x_1^2-\partial_+x_2\partial_-x_2 \,x_2^2\big)
-\frac{1}{8}\big((\partial_+ x_m)^2+(\partial_- x_m)^2\big)(x_1^2-x_2^2)
\\
&{}
+\frac{i}{2}(x_1^2-x_2^2)\big(\bar{\Theta}_+\Gamma^-\partial_+\Theta_++\bar{\Theta}_-\Gamma^-\partial_-\Theta_-\big)
+\frac{i}{2}\partial_-x^m\,\bar{\Theta}_-\Gamma^-\Gamma_m(x_1\Gamma^1-x_2\Gamma^2)\Theta_-
\nn\\
&{}
+\frac{i}{2}\partial_+x^m\,\bar{\Theta}_+\Gamma^-\Gamma_m(x_1\Gamma^1-x_2\Gamma^2)\Theta_+
-i\partial_-x^m\partial_+x^n\,\bar{\Theta}_-\Gamma_m\Gamma^-\Gamma^1\mathcal P_8\Gamma_n\Theta_+ +\mathcal{O}(\Theta^4)\,.\nn
\end{align}
We have omitted terms with four fermions because, as it turns out, they will not contribute below.\footnote{There are no quartic fermion terms consisting of only massive fermions.}

\section{Tree-Level Worldsheet Scattering Amplitudes}\label{sec:tree-level}

Equipped with the BMN Lagrangian to sixth order in fields we are now in position to perform explicit worldsheet computations. Our first task will be to determine all possible tree-level two to two scattering amplitudes.  S-matrix elements can be calculated at strong coupling by computing four-point functions on the string worldsheet. While the complexity of the problem increases dramatically with each loop order, the tree-level part is fairly easy to determine. 

Unfortunately, unlike the other examples of $AdS/CFT$ studied so far, there is no proposal for an exact S-matrix in the $AdS_2/CFT_1$ case to compare our results to. It would be very interesting to try to come up with such a proposal based on the symmetries of the problem. The symmetry preserved by the BMN vacuum in $AdS_2\times S^2\times T^6$ is a $\mathfrak u(1)$ central extension of the semi-direct sum of $\mathfrak u(1)$ and $\mathfrak{psu}(1|1)^2$, as can be seen by looking at the commutant of $P^+$ in the symmetry algebra in \cite{Sorokin:2011rr}. This is precisely half of the symmetry of $AdS_3\times S^3\times T^4$ (modulo central extensions) and roughly the same as that of $AdS_3\times S^3\times S^3\times S^1$. However the obvious way to construct an S-matrix with this symmetry by tensoring two $\mathfrak{su}(1|1)$ S-matrices \cite{Beisert:2005wm} does not seem to work in this case. Since we have not found a suitable exact S-matrix we will instead compare our massive tree-level scattering amplitudes to a truncation of the exact $\mathfrak{psu}(2,2|4)$ S-matrix of $AdS_5/CFT_4$; see \cite{Hoare:2013pma,Hoare:2013ida} for a similar discussion for the $AdS_3\times S^3\times T^4$ string. As we will show in appendix \ref{sec:ads3-truncation} the tree-level massive sector S-matrix exactly matches the one obtained from this truncation.

The S-matrix is normally divided into two factors, a matrix part which in principle may be determined from the underlying centrally extended symmetry algebra, and the dressing phase. We write this as 
\begin{equation}\label{eq:defn-Shat-Theta}
\mathbbm{S}_{ab}^{cd}(p_{1},p_{2})=\hat{\mathbbm{S}}_{ab}^{cd}(p_{1},p_{2})\, e^{i{\theta}(p_{1},p_{2})}
\end{equation}
and expand the dressing phase at strong coupling
\begin{equation}\label{eq:defn-theta(n)}
{\theta}=\sum_{n=0}^{\infty}h^{1-n}\theta^{(n)}
\end{equation}
defining terms $\theta^{(n)}$. These are order $h^{0}$ functions of the momenta in the regime 
when these are order 1 in the spin-chain normalisation. In
the BMN limit, where these momenta are very small, the one-loop phase
$\theta^{(1)}$ is order $1/h^{2}$ as a result of scaling like $p^{2}$. 
(That is the topic of section \ref{sec:one-loop}.)
Expanding $\mathbbm S$ to one-loop order, we arrange the terms as follows:
\bea 
\mathbbm S=\mathbbm 1 + i \frac{1}{h} \mathbbm{T}^{(0)}+i \frac{1}{h^2}\big[\mathbbm{T}^{(1)}+h^2 \theta^{(1)}\big]+\bigodiv{h^3}.
\eea 
Thus we define $\mathbbm{T}^{(0)}$ to include the effect of the AFS phase, but $\mathbbm{T}^{(1)}$ to be solely the result of the matrix part. %
We write this expansion in terms of the coupling $h$ of the integrable structure. This is related to the string theory's $\hat g$ by $h=2\hat g+\bigo{\hat g^{-2}}$, where the absence of corrections up to two loops was shown by \cite{Murugan:2012mf}.

As in section \ref{stringsection} we use a real representation for the massive bosons, and complex representation for the remaining bosonic and fermionic fields. We introduce creation an annihilation operators as follows:
\bea \nn
&& x^r=\frac{1}{\sqrt{2\pi}}\int \frac{dp}{\sqrt{2\omega_p}}\big( a_p^r e^{-i p\cdot \sigma} + \bar a_p^r e^{ip\cdot \sigma}\big),\qquad \chi^1_\pm=\frac{1}{\sqrt{2\pi}}\int dp \sqrt{\frac{p_\pm}{2 \omega_p}}\big(\chi_p^1 e^{- ip\cdot \sigma} \pm \bar \xi_p^1 e^{ip\cdot \sigma}\big), \\ \nn
&& y^i=\frac{1}{\sqrt{2\pi}}\int \frac{dp}{\sqrt{2|p|}}\big( b_p^i e^{-i p\cdot \sigma} + \bar c_p^i e^{ip\cdot \sigma}\big),\qquad \chi^i_\pm=\frac{1}{\sqrt{2\pi}}\int dp \sqrt{\frac{p_\pm}{2 |p|}}\big(\chi_p^i e^{- ip\cdot \sigma} \pm \bar \xi_p^i e^{ip\cdot \sigma}\big),
\eea
where $r=1,2,i=2,3,4$ and $\omega_p=\sqrt{1+p^2}$.  Our conventions are such that $\bar b_p^i$ and $\bar \chi_p^i$ creates a particle with positive i'th charge while $\bar c_p^i$ and $\bar \xi_p^i$ creates a corresponding anti-particle with negative charge.

\subsection{Massive to massive scattering}\label{sec:massive-massive}
We start with recording all massive to massive processes where we will write our results in terms of functions $\ell_i$ \eqref{eq:functions-ell}. Furthermore assuming that $p_1>p_2$ we find for two bosons\footnote{Here $\chi=\chi^1$ and $\xi=\xi^1$ throughout.}
\bea \nn
&& \mathbbm T^{(0)} \ket{a_{p_1}^1 a_{p_2}^1}=-\ell_1 \ket{a_{p_1}^1 a_{p_2}^1}-\ell_3\ket{\chi_{p_1} \xi_{p_2}}-\ell_3\ket{\xi_{p_1} \chi_{p_2}},
\\ \nn &&
\mathbbm T^{(0)} \ket{a_{p_1}^2 a_{p_2}^2}=\ell_1 \ket{a_{p_1}^2 a_{p_2}^2}+\ell_3\ket{\chi_{p_1} \xi_{p_2}}+\ell_3\ket{\xi_{p_1} \chi_{p_2}}, \\ \nn
&& \mathbbm{T}^{(0)} \ket{a_{p_1}^1 a_{p_2}^2}=\ell_2 \ket{a_{p_1}^1 a_{p_2}^2}-\ell_4\ket{\chi_{p_1} \xi_{p_2}}+\ell_4\ket{\xi_{p_1} \chi_{p_2}},
\\ \nn &&
\mathbbm{T}^{(0)} \ket{a_{p_1}^2 a_{p_2}^1}=-\ell_2 \ket{a_{p_1}^2 a_{p_2}^1}-\ell_4\ket{\chi_{p_1} \xi_{p_2}}+\ell_4\ket{\xi_{p_1} \chi_{p_2}} .
\eea
For one boson and one fermion:
\bea \nn
&& \mathbbm T^{(0)} \ket{a^1_{p_1} \chi_{p_2}}=-\ell_5 \ket{a_{p_1}^1 \chi_{p_2}}+i \ell_4\ket{\chi_{p_1} a^1_{p_2}}-i \ell_3 \ket{\chi_{p_1} a^2_{p_2}},\\ \nn
&&
\mathbbm T^{(0)} \ket{\chi_{p_1}a^1_{p_2} }=-\ell_6 \ket{\chi_{p_1}a_{p_2}^1 }+i \ell_4\ket{ a^1_{p_1}\chi_{p_2}}+i \ell_3 \ket{a^2_{p_1}\chi_{p_2}}, \\ \nn
&& \mathbbm T^{(0)} \ket{a^2_{p_1} \chi_{p_2}}=\ell_5 \ket{a_{p_1}^2 \chi_{p_2}}-i \ell_3\ket{\chi_{p_1} a^1_{p_2}}-i \ell_4 \ket{\chi_{p_1} a^2_{p_2}}, \\  \nn
&&
\mathbbm T^{(0)} \ket{\chi_{p_1}a^2_{p_2} }=\ell_6 \ket{\chi_{p_1}a_{p_2}^2 }+i \ell_3\ket{ a^1_{p_1}\chi_{p_2}}-i \ell_4 \ket{a^2_{p_1}\chi_{p_2}}, \\ \nn
&& \mathbbm T^{(0)} \ket{a^1_{p_1} \xi_{p_2}}=-\ell_5 \ket{a_{p_1}^1 \xi_{p_2}}+i \ell_4\ket{\xi_{p_1} a^1_{p_2}}+i \ell_3 \ket{\xi_{p_1} a^2_{p_2}}, \\ \nn
&&
\mathbbm T^{(0)} \ket{\xi_{p_1}a^1_{p_2} }=-\ell_6 \ket{\xi_{p_1}a_{p_2}^1 }+i \ell_4\ket{ a^1_{p_1}\xi_{p_2}}- i\ell_3 \ket{a^2_{p_1}\xi_{p_2}}, \\ \nn
&& \mathbbm T^{(0)} \ket{a^2_{p_1} \xi_{p_2}}=\ell_5 \ket{a_{p_1}^2 \xi_{p_2}}+i \ell_3\ket{\xi_{p_1} a^1_{p_2}}-i \ell_4 \ket{\xi_{p_1} a^2_{p_2}}, \\ \nn
&&
\mathbbm T^{(0)} \ket{\xi_{p_1}a^2_{p_2} }=\ell_6 \ket{\xi_{p_1}a_{p_2}^2 }-i \ell_3\ket{ a^1_{p_1}\xi_{p_2}}-i \ell_4 \ket{a^2_{p_1}\xi_{p_2}} .
\eea
And finally for two fermions:
\bea \nn
&& \mathbbm T^{(0)} \ket{\chi_{p_1}\chi_{p_2}}=0,\qquad \mathbbm T^{(0)} \ket{\chi_{p_1}\xi_{p_2}}=-\ell_3 \ket{a^1_{p_1} a^1_{p_2}}+\ell_3 \ket{a^2_{p_1} a^2_{p_2}}+\ell_4\ket{a^1_{p_1} a^2_{p_2}}+\ell_4 \ket{a^2_{p_1}a^1_{p_2}}, \\ \nn
&& \mathbbm T^{(0)} \ket{\xi_{p_1}\xi_{p_2}}=0,\qquad \, \,\mathbbm T^{(0)} \ket{\xi_{p_1}\chi_{p_2}}=-\ell_3 \ket{a^1_{p_1} a^1_{p_2}}+\ell_3 \ket{a^2_{p_1} a^2_{p_2}}-\ell_4\ket{a^1_{p_1} a^2_{p_2}}-\ell_4 \ket{a^2_{p_1}a^1_{p_2}} .
\eea
The coefficients $\ell_i$ are given by
\bea \label{eq:functions-ell}
&& \ell_1=\frac{1}{2\hat{g}}\frac{p_1^2+p_2^2}{\omega_2 p_1-\omega_1 p_2},\qquad \ell_2=\frac{1}{2\hat{g}}\frac{p_1^2-p_2^2}{\omega_2 p_1-\omega_1 p_2}, \\ \nn
&&
\ell_3=\frac{1}{4\hat{g}}\frac{p_1 p_2}{p_1+p_2}\big(\sqrt{(\omega_1+p_1)(\omega_2+p_2)}-\sqrt{(\omega_1-p_1)(\omega_2-p_2)}\big),\\ \nn
&&
\ell_4=\frac{i}{4\hat{g}}\frac{p_1 p_2}{p_1-p_2}\big(\sqrt{(\omega_1+p_1)(\omega_2+p_2)}+\sqrt{(\omega_1-p_1)(\omega_2-p_2)}\big), \\ \nn
&& \ell_5=\frac{1}{2\hat{g}}\frac{p_2^2}{\omega_2 p_1-\omega_1 p_2},\qquad
\ell_6=\frac{1}{2\hat{g}}\frac{p_1^2}{\omega_2 p_1-\omega_1 p_2}\,.
\eea
The above amplitudes can be reproduced by truncating the exact {$AdS_3\times S^3\times T^4$} S-matrix of \cite{Borsato:2013qpa}. However, this truncation is only valid at tree-level which we explicitly demonstrate in appendix \ref{sec:ads3-truncation}.

\subsection{Including massless modes}
What happens if we scatter particles of different mass? Starting with the simplest case of two massive ingoing and two massless outgoing particles, the allowed possible processes are
\bea \nn 
\ket{a^s a^r}\rightarrow \ket{b^i c^i}\quad \textrm{or} \quad \ket{a^s a^r}\rightarrow \ket{\chi^i \xi^i},\qquad i=2,3,4
\eea 
where energy and momentum conservation implies
\bea \nn
p_3=\frac{1}{2}\big(p_1+p_2\big)\pm\frac{1}{2}\big(\omega_1+\omega_2\big),
\qquad
p_4=\frac{1}{2}\big(p_1+p_2\big)\mp\frac{1}{2}\big(\omega_1+\omega_2\big)
\eea
hence $p_3$ and $p_4$ have opposite sign. While most amplitudes vanish trivially, this condition forces the remaining to be zero once the energy momentum constraints are satisfied. Hence, all massive to massless processes are zero,
\[
\mathbbm{T}^{(0)} \ket{a^s a^r}=0.
\] 
For the case of incoming particles of different mass the situation is more complicated. For simplicity we will only consider the case where the incoming state is purely bosonic. Furthermore, the energy momentum constraints have two solutions: one where the particles of different mass keep their momenta and one where there is a non-trivial exchange. However, integrability implies that the particle can, at most, exchange their momenta and since classical integrability was proved in \cite{Sorokin:2011rr} we will assume this to be the case. Computing the amplitudes gives
\[
\mathbbm{T}^{(0)} \ket{a^r_{p_1} b^{i}_{p_2}}=(-1)^r\gamma_1\ket{a^r_{p_1} b^i_{p_2}},\qquad 
\mathbbm{T}^{(0)} \ket{a^r_{p_1} c^{i}_{p_2}}=(-1)^r\gamma_1\ket{a^r_{p_1} c^i_{p_2}}
\] 
where the coefficients are given by\footnote{In the amplitudes we have also included the Jacobian and external leg factors which combined gives an overall contribution
\bea \nn 
\frac{1}{4}\frac{1}{|p_2|\big(p_1-\omega_1 \text{sign}(p_2)\big)}.
\eea
}
\[
\gamma_1=\frac{i}{2}\frac{|p_2|}{p_1-\omega_1 \text{sign}(p_2)}.
\] 
Finally we want to study processes involving only massless particles. Again we will restrict to cases where we have two incoming bosonic particles. From (\ref{eq:quarticL}) it's clear that we will not have any purely bosonic processes since there are no quartic boson interaction terms. However, for an outgoing state of two fermions we find
\[
\mathbbm{T}^{(0)}\ket{b^i_{p_1} c^j_{p_2}}=
\begin{cases}
0,  
& \text{sign}(p_1)= \text{sign}(p_2) \\
-\frac{1}{2}i^{\delta_{ij}}  \sqrt{|p_1 p_2|}  \ket{\chi_{p_1}^i \xi^j_{p_2}}, 
& \text{sign}(p_1)\neq \text{sign}(p_2).
\end{cases}
\] 
The first case can be excluded on physical grounds since if the particles have momenta of the same sign, they would never meet. \subsection{SU(2) Bethe equations}
Based on the $\frac{PSU(1,1|2)}{SO(1,1)\times U(1)}$ coset a set of Bethe equations were proposed in \cite{Sorokin:2011rr}. Restricted to the SU(2) sector they take the form
\bea \label{eq:diagonal-amplitude}
\Big(\frac{x^+_k}{x^-_k}\Big)^L=\prod_{j\neq k}^M\frac{x^+_k-x^-_j}{x^-_k-x^+_j}\frac{1-\frac{1}{x^+_kx^-_j}}{1-\frac{1}{x^-_k x^+_j}} e^{i{\theta}(x_k,x_j)}
\eea
where the phase, written in the string frame, is given by
\bea \nn
e^{i{\theta}(x_1,x_2)}=e^{-\frac{i}{h}\big(\omega_1 p_2-\omega_2 p_1\big)}\Big(\frac{x^-_1\,x^+_2}{x^+_1\,x^-_1}\Big)^2\sigma_\text{AFS}^4+\mathcal{O}(h^{-2}).
\eea
Focusing on $M=2$, the right hand side of the equations should reproduce the diagonal tree-level amplitude $\ell_1$. A quick calculation gives
\bea
1-\frac{i}{h}\frac{p_1^2+p_2^2}{\omega_2p_1-\omega_1 p_2}+\mathcal{O}(h^{-2})
\eea
which, using (\ref{eq:coupling-relations}), is in nice agreement with $\ell_1$ in (\ref{eq:functions-ell}). This confirms that the AFS phase indeed enters with a power of four, see also \cite{Murugan:2012mf}.

\subsection{Probing the one-loop sector with tree-level amplitudes}\label{sec:conn-one-two}
It is a well known fact in quantum field theory that higher order amplitudes partially can be determined via lower loop components. This is an especially powerful tool at the leading one-loop level where almost the full amplitude can be determined via generalized unitarity and the optical theorem. A generic one-loop amplitude in two dimensions, with fourth- and sixth-order interactions, can be written as
\bea
\mathbbm{T}^{(1)}=i f_1(p_1,p_2)\log\frac{p_1}{p_2}+i f_2(p_1,p_2) + f_3(p_1,p_2)
\eea
where $f_i$ are real. The first function is completely determined through $s$ and $u$-channel contributions while the last two generally have contributions from all types of diagrams. Generalized unitarity allows us to completely determine $f_1$, while the optical theorem will give us $f_3$.

\subsubsection*{Generalized unitarity}
Here we will closely follow \cite{Engelund:2013fja}. We start by introducing the notation,
\bea \nn
(\mathbbm{T}^{(0)})_{CD'}^{AB'}=\bra{D' C} \mathbbm{T}^{(0)} \ket{A B'}
\eea
where the primed index denotes an excitation with momentum $p_2$. We furthermore introduce $3$ and $4$ labels for $\chi$ and $\xi$ particles. Using this, the $s$ and $u$-channel cuts are contained in the following two expressions
\bea \nn
(C_s)^{CD'}_{AB'}=J i^2 (i\mathbbm T^{(0)})^{CD'}_{EF'} (i\mathbbm T^{(0)})_{AB'}^{EF'},\qquad (C_u)^{CD'}_{AB'}=i^2(-1)^{(|B|+|F|)(|D|+|F|)}J(i\mathbbm T^{(0)})^{CF'}_{EB'}(i\mathbbm T^{(0)})^{ED'}_{AF'}
\eea
where J is the relativistic Jacobian. The first function $f_1$ can be determined from
\bea \nn
J\big(C_s-C_u\big)\,.
\eea
Focusing on the $x_1 x_1\rightarrow x_1 x_1$ and $x_1 x_2\rightarrow x_1 x_2$ amplitudes we find
\begin{align}
\label{eq:generalized-unitarity-amp}
 x_1 x_1\rightarrow x_1 x_1:\qquad & (\mathbbm T^{(0)})^{11'}_{EF'}(\mathbbm T^{(0)})^{EF'}_{11'}- (-1)^{|F|}(\mathbbm T^{(0)})^{1F'}_{E1'}(\mathbbm T^{(0)})^{E1'}_{1F'}
\\ \nn
 &=(\ell_1^2+2\ell_3^2)-(\ell_1^2+2\ell_4^2)=\frac{1}{2}\Big(\frac{p_1p_2}{\omega_1 p_2 - \omega_2 p_1}\Big)^2 \vec p_1 \cdot \vec p_2,\\ \nn
 x_1 x_2\rightarrow x_1 x_2:\qquad &
 (\mathbbm T^{(0)})^{12'}_{EF'}(\mathbbm T^{(0)})^{EF'}_{12'} -(-1)^{|F|}(\mathbbm T^{(0)})^{1F'}_{E2'}(\mathbbm T^{(0)})^{E2'}_{1F'}
\\ \nn
&=(\ell_2^2-2\ell_4^2)-(\ell_2^2-2\ell_3^2) = \frac{1}{2}\Big(\frac{p_1p_2}{\omega_1 p_2 - \omega_2 p_1}\Big)^2 \vec p_1 \cdot \vec p_2\,.
\end{align}
Hence the two amplitudes give identical results, something which is expected since, at one-loop, the imaginary part of the amplitude should come solely from the scattering phase which acts diagonally.

\subsubsection{Optical theorem}
A well known consequence of unitarity of the S-matrix is that the real part of a one-loop amplitude can be obtained via tree-level processes,
\bea \nn
\textrm{Im} \mathbbm (T^{(1)})_{AB'}^{CD'}=(\mathbbm{T}^{(0)})^{CD'}_{EF'}((\mathbbm{T}^{(0)})^\dagger)^{EF'}_{AB'}\,.
\eea
For the two processes above, this quantity can be read off from (\ref{eq:generalized-unitarity-amp}),
\bea
\label{eq:optical-theorem-amps}
&&\negthickspace\negthickspace  x_1 x_1\rightarrow x_1 x_1: \\ \nn
&&\big(\ell_1^2+2\ell_3^2\big)=\frac{\big((p^1_-)^2+(p_-^2)^2+(p_-^1p_-^2)^2\big((p_-^1)^2+(p_-^2)^2-4\big)\big)^2}{16(p_-^1p_-^2)^2\big((p_-^2)^2-(p_-^1)^2\big)^2}
+\frac{\big(1-(p_-^1)^2\big)^2\big(1-(p_-^2)^2\big)^2}{32(p_-^1p_-^2)\big(p_-^1+p_-^2\big)^2} \\ \nn
&&\negthickspace \negthickspace  x_1 x_2\rightarrow x_1 x_2: \\ \nn
&& \big(\ell_2^2+2|\ell_4|^2\big)=
\frac{1}{4}\Big(\frac{p_1p_2}{\omega_1 p_2-\omega_2 p_1}\Big)^2\Big(\frac{(p_1^2-p_2^2)^2}{(p_1 p_2)^2}+1-p_1p_2 +\omega_1 \omega_2\Big)
\eea
where we choose to write the first amplitude in terms of right moving momenta. We do this to facilitate a simple comparison with later computations performed in the near-flat-space limit.

\section{One-Loop Scattering Amplitudes}\label{sec:one-loop}

In this section we will calculate the one-loop amplitudes for scattering involving the two massive bosons $x_1$ and $x_2$. Some of these amplitudes are divergent in the near-BMN limit, including the diagonal process $x_1 x_1\rightarrow x_1 x_1$. This can probably be traced back to the gauge-fixing which becomes more involved at the quantum level. For divergent amplitudes we will isolate the finite pieces of the $s$ and $u$-channel contributions, which still will allow us to determine most terms of the one-loop dressing phase. The non-diagonal processes $x_1 x_2\rightarrow x_1 x_2$ and $x_1 x_1\rightarrow x_2 x_2$ are, however, finite and these amplitudes we determine completely.

A fairly involved computation gives
\bea
\label{eq:BMN-amplitude}
&& \ell_1^{(1)}=\frac{1}{4\pi\hat g^2}\Big(\frac{p_1\, p_2 }{\omega_1 p_2-\omega_2 p_1}\Big)^2\Big[i\big(\omega_2 p_1-\omega_1 p_2- \vec p_1\cdot \vec p_2\log{p^2_- \over p^1_-}\big)+\dots\Big], \\ \nn
&&  \ell_2^{(1)}=\frac{1}{4\pi\hat g^2}\Big(\frac{p_1\, p_2 }{\omega_1 p_2-\omega_2 p_1}\Big)^2\Big[i\big(\omega_2 p_1-\omega_1 p_2- \vec p_1\cdot \vec p_2\log{p^2_- \over p^1_-}\big)-\frac{\pi}{2}\Big(\frac{(p_1^2-p_2^2)^2}{(p_1p_2)^2}+1-p_1p_2+\omega_1\omega_2\Big)\Big]
\eea
where $\ell_1^{(1)}$ is the one-loop contribution to diagonal $x_1 x_1\rightarrow x_1 x_1$ scattering and $\ell_2^{(1)}$ is the corresponding $x_1 x_2\rightarrow x_1 x_2$ amplitude.  The UV finiteness of the second amplitude comes from a delicate cancellation of $1/\epsilon$ terms between four and six order vertices.

For both amplitudes $\ell^{(1)}_i$ we see that the transcendental part is completely reproduced via the tree-level amplitudes in (\ref{eq:generalized-unitarity-amp}). Furthermore we also notice that the real part of the second amplitude $\ell^{(1)}_2$  matches with the terms obtained through the optical theorem in (\ref{eq:optical-theorem-amps}).

Using the standard form of the  \5 Hern\'{a}ndez--L\'{o}pez phase\footnote{We denote generic one-loop phases by $e^{i \theta^{(1)}}$, and write $\sigma_\text{HL}^2=e^{i 2 \theta_\text{HL}}$ for this particular phase from $AdS_5$.} in the BMN limit \cite{Klose:2007wq, Engelund:2013fja}
\bea
\label{eq:hl-phase}
&&2\,\theta_\text{HL}=\frac{1}{\pi h^2}\Big(\frac{p_1\, p_2 }{\omega_1 p_2-\omega_2 p_1}\Big)^2\Big[\omega_2 p_1-\omega_1 p_2- \vec p_1\cdot \vec p_2\log{p^2_- \over p^1_-}\Big]
\eea
we see that the imaginary part of \eqref{eq:BMN-amplitude} is identical to this given $h=2\hat{g}$. 
Thus up to one-loop order the complete dressing phase takes the form
\bea
e^{i{\theta}} = \sigma^4_\text{AFS}\sigma^2_\text{HL}+\mathcal{O}(h^{-3})
\eea
where we point to \eqref{eq:diagonal-amplitude} for the classical comparison. %

At the one loop level there are   additional processes not present at the classical level. For example, $x_1 x_1\rightarrow x_2 x_2$ is one such process and since it turns out to be UV finite we record it here,
\bea
&& x_1 x_1\rightarrow x_2 x_2: \\ \nn
&& \frac{1}{8\pi}\frac{1}{\omega_2 p_1 - \omega_1 p_2}\Big[\frac{i\big(8+7(p_1^2+p_2^2)+6\,p_1 p_2+8\,\omega_1 \omega_2\big)}{6}+\frac{\pi(p_1p_2)^2\big(\omega_1-\omega_2\big)}{p_1+p_2}\Big].
\eea
By unitarity, the opposite amplitude $x_2 x_2\rightarrow x_1 x_1$ gives the same answer.

\subsection{Near-flat-space amplitudes and truncation}

As we described above, the diagonal amplitude fails to be UV finite. Nevertheless we can still probe parts of the amplitude by using the near-flat-space (NFS) limit of\cite{Maldacena:2006rv}.\footnote{See \cite{Abbott:2011xp,Sundin:2013ypa} for related work in $AdS_3$ and $AdS_4$.} This limit can be thought of as a subsector of the BMN string %
in which we keep only the leading terms under a large worldsheet Lorentz boost: 
\bea
p_\pm \rightarrow \hat g^{\mp \frac{1}{2}}\,p_\pm,\qquad \chi_\pm \rightarrow \hat g^{\mp \frac{1}{4}}\chi_\pm\,.
\eea
Since the right-moving sector is boosted, some of the subleading term will be pushed to zeroth order in the large coupling expansion. Keeping only these terms a much simpler Lagrangian, still with non-trivial interactions, emerges
\bea
\nn
\mathcal{L}_\text{BMN}=\mathcal{L}_2+\frac{1}{\hat g}\mathcal{L}_4(\partial_\pm x,\chi_\pm)+\mathcal{O}(\hat g^{-2})\qquad\Longrightarrow\qquad\mathcal{L}_{nfs}=\mathcal{L}_2+\mathcal{L}'_4(\partial_- x,\chi_-)+\mathcal{O}(\hat g^{-1})\,.
\eea
Furthermore, in loop integrals we only have powers of right-moving momenta in the numerator and since these are zero the resulting theory is manifestly UV finite. Also, by simple power counting, it is easy to see that only the quartic vertices contribute. This allows us to circumvent the problem with higher-order gauge fixing encountered for the BMN string and therefore we can determine the complete (NFS) form of the one-loop amplitude. A quick calculation gives
\bea
\label{eq:nfs-1loop}
&& (\ell_1^{(1)})_{NFS}=
\frac{1}{32\pi}\frac{\big(p_-^1 p_-^2\big)^2}{(p_-^2)^2-(p^1_-)^2}\Big[\color{red} \gamma \color{black} i p_-^1 p_-^2 -i\frac{p_-^1 p_-^2\big((p_-^2)^2+(p_-^1)^2\big)}{(p_-^2)^2-(p_-^1)^2}\log \frac{p_-^2}{p_-^1} \\ \nn
&&-\frac{\pi}{2}\frac{\big(2 (p_-^1)^4+(p_-^1)^3 (p_-^2)+2(p_-^1)^2(p_-^2)^2+p_-^1 (p_-^2)^3+2(p_-^2)^4\big)}{(p_-^2)^2-(p_-^1)^2}\Big]\,.
\eea
This amplitude has been computed with and without the massless modes (i.e. full string action and supercoset model respectively). The difference between the two is encoded in the parameter {\color{red}$\gamma$},
\bea \nn
&& AdS_2\times S^2:\qquad \qquad\,\,\color{red}\gamma\color{black}=1 \\ \nn
&& AdS_2\times S^2\times T^6:\qquad \color{red}\gamma\color{black}=4
\eea
and thus the only change is an overall numerical factor in front of a single term (which arises from a $t$-channel type diagram). We also note that, in both cases, the real part is reproduced by taking the NFS-limit of (\ref{eq:optical-theorem-amps}). Furthermore, for the ten-dimensional case, the imaginary part above is identical to the one-loop corrected NFS amplitude in \5 \cite{Klose:2007wq}.

For the non-diagonal amplitude, which we can calculate in the BMN limit, the difference between the full string and the supercoset model is more involved
\bea \nn
&& \ell^{(1)}_2=\frac{i}{32\pi}\frac{1}{\omega_2 p_1-\omega_1p_2}\Big(3+6\big(p_1^2+p_2^2\big)+12\,p_1 p_2\,\omega_1\omega_2+20\,p_1^2p_2^2\Big)+\dots
\eea
where the dots denote the last three terms in (\ref{eq:BMN-amplitude}), i.e. the transcendental terms and real factors. While this amplitude is still UV finite, we see that it changed to a structurally more complicated form. Naturally, since the $s$ and $u$-channel contributions can be written in terms of tree-level amplitudes, it is clear that any changes observed with the massless modes turned off is in $t$-channel and six-vertex terms. This follows from the fact that, at the classical level, the equations of motion for the massless states allow for the trivial zero solution. Thus, any terms of a one-loop amplitude which can be expressed in terms of tree-level amplitudes must be invariant under the truncation to the massive sector. 

\section{The Semiclassical Dressing Phase}\label{sec:semiclass-HL}

There is another way to calculate quantum corrections to the dressing
phase, independently of the above worldsheet calculation. This is
the semiclassical calculation following \cite{Chen:2007vs} in $AdS_{5}$,
where this was used to check \cite{Hernandez:2006tk}.%
\footnote{A closely related calculation was performed by \cite{Gromov:2007cd}.%
} Similar calculations were performed in $AdS_{3}$ by \cite{David:2010yg,Beccaria:2012kb,Abbott:2013mpa},
where it was seen that the phase is different to that for $AdS_{5}$.
The calculation here follows \cite{Abbott:2013mpa} closely, and thus
we do not show much detail. 

\newcommand{\smalldynkinT}[3]{
\begin{tikzpicture}[scale=0.6]
\draw (1,0) -- (3,0);
\draw [fill=#1] (1,0) circle (2mm); 
\draw [fill=#2] (2,0) circle (2mm); 
\draw [fill=#3] (3,0) circle (2mm); 

\draw [thin] (8mm,0) -- (12mm,0);
\draw [thin] (1,-2mm) -- (1,2mm);

\draw [thin] (28mm,0) -- (32mm,0);
\draw [thin] (3,-2mm) -- (3,2mm);

\end{tikzpicture}
} 
\newcommand{\colourone}{midblue}
\newcommand{\smallcirc}[1]{\begin{tikzpicture}[scale=0.6] \draw [fill=#1] (0,0) circle (2mm); \end{tikzpicture}}

The description of $AdS_{2}\times S^{2}$ as the coset $PSU(1,1|2)/U(1)\times U(1)$
is one of the cases considered by \cite{Zarembo:2010yz}. The Dynkin
diagram is \smalldynkinT{white}{white}{white} thus the Cartan matrix
is the same as that for the left sector of $AdS_{3}\times S^{3}$
alone: 
\[
A=\left[\begin{array}{ccc}
0 & -1 & 0\\
-1 & 2 & -1\\
0 & -1 & 0
\end{array}\right].
\]
There are three quasimomenta $p_{\ell}(z)$, $\ell=1,2,3$, and the
BMN vacuum is given by the same $\kappa=(1,0,1)$ as for $AdS_{3}\times S^{3}$
i..e 
\[
p_{1}^{\text{vac}}(z)=p_{3}^{\text{vac}}(z)=\frac{\Delta}{2g}\frac{z}{z^{2}-1},\qquad p_{2}^{\text{vac}}(z)=0.
\]
The list of possible modes is also like the left sector of $AdS_{3}\times S^{3}$
(see for instance \cite{Abbott:2012dd}'s appendix), which in turn
is a truncation of the list in $AdS_{5}\times S^{5}$: \begin{equation} 
\begin{tabular}{lcr|cl}
\hline 
Bosons ($y_{1}$): & \smalldynkinT{white}{\colourone}{white} & $r=S$$\vphantom{\frac{1^{1}}{1_{1}}}$ & $2\pi n(x)=\negthickspace\negthickspace\negthickspace$ & $-p_{1}(z)+2p_{2}(z)-p_{3}(z)$\tabularnewline
 & \smalldynkinT{\colourone}{\colourone}{\colourone} & $A$$\vphantom{\frac{1^{1}}{1_{1}}}$ &  & $-p_{1}(z)-p_{3}(z)$\tabularnewline
\hline 
Fermions ($\chi_{1}$): & \smalldynkinT{\colourone}{\colourone}{white} & $F$$\vphantom{\frac{1^{1}}{1_{1}}}$ &  & $-p_{1}(z)+p_{2}(z)-p_{3}(z)$\tabularnewline
 & \smalldynkinT{white}{\colourone}{\colourone} & $F'$$\vphantom{\frac{1^{1}}{1_{1}}}$ &  & $-p_{1}(z)+p_{2}(z)-p_{3}(z)$\tabularnewline
\hline 
\end{tabular} \label{eq:modes-colouring-in} \end{equation}  

However the action of the $\mathbb{Z}_{4}$ symmetry is different
here. This is encoded in the matrix $S$ \cite{Zarembo:2010yz} 
\[
S=\small\left[\begin{array}{ccc}
 &  & -1\\
 & -1\\
-1
\end{array}\right]
\]
which leads to inversion conditions $p_{\ell}(\tfrac{1}{z})=S_{\ell m}p_{m}(z)$
i.e. 
\[
p_{1}(z)=-p_{3}(\tfrac{1}{z}),\qquad p_{2}(z)=-p_{2}(\tfrac{1}{z}).
\]

We expect only one possible polarisation of giant magnon in $S^{2}$,
and this is given by a single log cut resolvent $G(z)$ on sheet $\ell=2$,
as for the mode labelled ``$S$'' in \eqref{eq:modes-colouring-in}.
The inversion symmetry then demands $-G(\tfrac{1}{z})$ on the same
sheet, and so we have 
\[
p_{1}(z)=p_{3}(z)=\frac{\Delta}{2g}\frac{z}{z^{2}-1},\qquad p_{2}(z)=G(z)-G(\tfrac{1}{z})
\]
where 
\begin{equation}\label{eq:Gmag}
G(z)=-i\log\frac{z-\Xp}{z-\Xm}=-\sum_{n=0}^{\infty}Q_{n+1}z^{n}.
\end{equation}
This defines the charges 
\[
Q_{n+1}=\frac{i}{n}\Big(\frac{1}{\Xp^{n}}-\frac{1}{\Xm^{n}}\Big),\qquad n\geq1.
\]
In \cite{Abbott:2013mpa} it was necessary to add a constant twist
$-p/2$ to this resolvent, to account for the fact that a single giant
magnon is not a closed string \cite{Gromov:2008ie,Gromov:2007ky}.
Here it is clear that this will cancel out of $p_{2}(z)$ and thus
have no effect. 

We are now interested in modes of this solution, and in particular
their phase shifts 
\[
\delta_{rS}(z,\Xpm)=2\pi n_{r}-2\pi n_{r}^{\mathrm{vac}}
\]
The giant magnon above leads to the first colum in the following table:
\begin{equation} %
\begin{tabular}{c|c|c|c|}
 & $\delta_{rS}(z,\Xpm)$ & $\delta_{rA}(z,\Xpm)$ & $\delta_{rF}(z,\Xpm)=\delta_{rF'}(z,\Xpm)$\tabularnewline
\hline 
$\negthickspace\negthickspace\negthickspace r=S$ & $2[G(z)-G(\tfrac{1}{z})]$ & 0 & $G(z)-G(\tfrac{1}{z})$\tabularnewline
$A$ & 0 & $-2[G(z)-G(\tfrac{1}{z})]$ & $-G(z)+G(\tfrac{1}{z})$\tabularnewline
\hline 
$F$ & $G(z)-G(\tfrac{1}{z})$ & $-G(z)+G(\tfrac{1}{z})$ & 0\tabularnewline
$F'$ & $G(z)-G(\tfrac{1}{z})$ & $-G(z)+G(\tfrac{1}{z})$ & 0\tabularnewline
\hline 
\end{tabular}\label{eq:table-of-delta}\end{equation} 

Then we compute the correction \cite{Abbott:2013mpa,Chen:2007vs}
\begin{align}\label{eq:delta-delta}
\theta^{(1)}_{12}(p_{1},p_{2}) & =\frac{1}{4\pi}\int_{-1}^{1}dz\sum_{r}(-1)^{F_{r}}\left[\frac{\partial\delta_{r1}(z,e^{\pm ip_{1}/2})}{\partial z}\delta_{r2}(z,e^{\pm ip_{2}/2})\right]\\
 & =\frac{2}{4\pi}\sum_{\substack{r,s\geq2\\
r+s\text{ odd}
}
}c_{r,s}Q_{r}(p_{1})Q_{s}(p_{2})\nn
\end{align}
finding  the same coefficients as in \cite{Hernandez:2006tk}:
\[
c_{r,s}=-8\frac{(r-1)(s-1)}{(r+s-2)(s-r)},\quad r+s\text{ odd},\; r,s\geq2 .
\]
This is exactly the same result as in $AdS_{5}\times S^{5}$, including
the factors of 2. 

We can similarly compute the correction for any other pair of massive states, and they are all the same. The classical solutions being scattered are no longer  giant magnons but they can still be constructed using the resolvent \eqref{eq:Gmag}, on the pattern of the modes shown in \eqref{eq:modes-colouring-in}. The resulting phase shifts $\delta_r(z)$ are shown in \eqref{eq:table-of-delta}, and clearly give the same result in \eqref{eq:delta-delta}.

\section{Conclusions}\label{sec:Conclusions}

In this paper we have examined worldsheet scattering processes for
strings in $AdS_{2}\times S^{2}\times T^{6}$. The resulting magnon S-matrix
is the central object in the integrable description of AdS/CFT.

The main result  is that the dressing phase is like that of
the original $AdS_{5}$ integrable theory, at least to one loop. This
is interesting in light of the fact that the $AdS_{3}$ dressing phase
is not the same \cite{David:2010yg,Beccaria:2012kb,Borsato:2013hoa,Abbott:2013mpa},
which rules out the idea that the BES phase is truly universal. Let us summarise
this by writing the $AdS_5$ phase as
\[
\sigma^2=e^{2i\theta_{\text{BES}}},\qquad\theta_{\text{BES}}(h)=h\theta_{\text{AFS}}+\theta_{\text{HL}}+\bigodiv{h},\qquad h=\frac{\sqrt{\lambda}}{2\pi}.
\]
Then the dressing phases in the most famous integrable superstring
models are as follows, writing the exact phases alongside their strong-coupling
expansions:
\begin{equation}\label{eq:list-of-Theta}\begin{aligned}
\qquad\qquad AdS_{5}\times S^{5}: & \vphantom{\frac{1}{1_{1}^{1}}} & \theta=\; 
& 2 h \theta_{\text{AFS}}+2\theta_{\text{HL}}+\bigo{\tfrac{1}{h}}\negthickspace\negthickspace & \negthickspace\negthickspace & =2\theta_{\text{BES}}(h)\qquad\qquad\\
AdS_{4}\times CP^{3}: &  &  
& h\theta_{\text{AFS}}+\theta_{\text{HL}}+\ldots & \negthickspace\negthickspace & =\theta_{\text{BES}}(h)\\
AdS_{3}\times S^{3}\times T^{4}: &  &  
& 2h\theta_{\text{AFS}}+\begin{cases}
2\theta_{LL}+\ldots\\
2\tilde{\theta}_{LR}+\ldots
\end{cases} & \negthickspace\negthickspace & =\begin{cases}
2\theta_{\text{BOSST}}\\
2\tilde{\theta}_{\text{BOSST}}
\end{cases}\\
AdS_{2}\times S^{2}\times T^{6}: &  &  
& 4 h \theta_{\text{AFS}}+2\theta_{\text{HL}}+\ldots & \negthickspace\negthickspace & =2\theta_{\text{BES}}(2h).
\end{aligned}\end{equation}
That the classical phase was $\sigma_{\mathrm{AFS}}^{4}$ was noted
in \cite{Cagnazzo:2011at}, and this is consistent with our tree-level 
scattering amplitudes. The fact that at one loop we see  $\sigma_{\mathrm{HL}}^{2}$
(rather than $\sigma_{\mathrm{HL}}^{4}$) is not inconsistent with the exact phase 
being $\theta_{\text{BES}}$, so long as it is scaled as shown. 

In all of this there are two coupling constants: 
the Bethe coupling $h$ is what appears in the integrable
structure, while $\hat{g}=\sqrt{\lambda}/4\pi$ is (one half) the
effective string tension, and is the relevant coupling for the worldsheet
string theory. These are related in the present case by 
\begin{equation}\label{eq:coupling-relations}
h=2\hat g+  \bigodiv{ \hat g^{2} }
\end{equation}
where the absence of corrections up to two loops was shown by \cite{Murugan:2012mf} in the near-flat-space limit (and up to one loop in the BMN limit). Of the cases in the list \eqref{eq:list-of-Theta} above, only the $AdS_4$ case has corrections to this relationship at one loop.%
\footnote{In $AdS_4 \times CP^3$ the relation is $h=\smash{\sqrt{\lambda/2}}+c+\ldots$ (with $\lambda=N/k$ giving $R^2/\alpha'=2^{5/2}\pi\sqrt\lambda$). The constant $c$ depends on the cutoff prescription used \cite{McLoughlin:2008he,Abbott:2010yb,Abbott:2011xp}. The case of $AdS_3\times S^3 \times S^3 \times S^1$ appears to be quite similar, at least at $\alpha=1/2$ \cite{Sundin:2012gc,Abbott:2012dd,Beccaria:2012kb}.} 
Most papers on $AdS_{5}$ write only $\sqrt \lambda$,  building in the fact that this coupling, also known as the interpolating function, is $h(\lambda)=\sqrt{\lambda}/2\pi$ to all orders \cite{Hofman:2006xt,Berenstein:2009qd}.

The relation to the $AdS_{3}$ dressing phase is that we have the
sum of the left-left and the left-right phases: %
\[
\theta_{\text{HL}}=\theta_{LL}+\tilde{\theta}_{LR} .
\]
While the phases we call $\theta_{\text{BOSST}}$ and $\tilde{\theta}_{\text{BOSST}}$
are thought to be exact \cite{Borsato:2013hoa}, there have so far
been no tests beyond one loop. For this reason, and the scaling  $\theta_\text{BES}(2h)$ 
needed in \eqref{eq:list-of-Theta}, a two-loop check would be very interesting here. 
Two-loop corrections to scattering in $AdS_{5}$'s near-flat-space limit 
were calculated in \cite{Klose:2007rz}.

We computed the one-loop dressing phase in two different ways: a worldsheet
calculation (drawing Feynman diagrams for the near-BMN action) and
a semiclassical calculation (based on soliton energy corrections).
The former uses the full Green--Schwarz action and in particular includes all
the massless modes, while the latter uses only the coset model, thus
omitting the torus directions completely. We see perfect agreement
between these two calculations. 

This agreement is interesting in light of the fact that deleting the
massless modes from the worldsheet calculation (where they appear on internal lines of Feynman diagrams) changes the one-loop correction. 
This restriction is classically a consistent truncation, but this does not guarantee
that it is allowed beyond tree level. The scattering amplitudes can
change in ways which are not physical, but note that similarly omitting massless
modes from the two-point functions in \cite{Murugan:2012mf}\footnote{See equation (4.12) in \cite{Murugan:2012mf}, and discussion below.} changed
a clearly physical two-loop correction to the mass.

It would be interesting to understand this issue of whether the massless modes decouple more thoroughly. As mentioned above it would also be of interest to see whether the two-loop dressing phase is what is predicted by the scaling suggested here.

\subsubsection{Acknowledgements}

MCA is supported by a UCT URC postdoctoral fellowship. PS acknowledges the support of a Claude Leon postdoctoral grant. JM acknowledges support from the NRF of South Africa under the HCDE and IPRR programs. The research of LW is supported in part by NSF grant PHY-0906222.

\appendix

\section{Tree-level S-matrix by Truncation}
\label{sec:ads3-truncation}
Consider the truncation of the tree-level $AdS_3\times S^3\times T^4$ S-matrix in eq. (4.1) of \cite{Hoare:2013pma}, which is itself obtained as a truncation of the $AdS_5\times S^5$ S-matrix \cite{Klose:2006zd}, to the ''real'' fields
\begin{equation}
a^1=\frac12(z_++z_-)\,,\quad a^2=\frac12(y_++y_-)\,,\quad\xi=\frac12(i\zeta_+-\chi_-)\,,\quad\chi=\frac12(i\zeta_-+\chi_+)\,.
\end{equation}
The truncated S-matrix becomes:
\subsubsection*{Boson-Boson}
\begin{eqnarray}
\mathbbm T|a^1a^1\rangle&=&\frac12(-l_1-l_2+2c)|a^1a^1\rangle+\frac12 l_4|\chi\xi\rangle+\frac12 l_4|\xi\chi\rangle
\nonumber\\
\mathbbm T|a^2a^2\rangle&=&\frac12(l_1+l_2+2c)|a^2a^2\rangle-\frac12 l_4|\chi\xi\rangle-\frac12 l_4|\xi\chi\rangle
\nonumber\\
\mathbbm T|a^1a^2\rangle&=&(-l_3+c)|a^1a^2\rangle+\frac{i}{2}l_5|\chi\xi\rangle-\frac{i}{2}l_5|\xi\chi\rangle
\nonumber\\
\mathbbm T|a^2a^1\rangle&=&(l_3+c)|a^2a^1\rangle+\frac{i}{2}l_5|\chi\xi\rangle-\frac{i}{2}l_5|\xi\chi\rangle
\end{eqnarray}
\subsubsection*{Boson-Fermion}
\begin{eqnarray}
\mathbbm T|a^1\xi\rangle&=&\frac12(-l_6-l_7+2c)|a^1\xi\rangle+\frac12 l_5|\xi a^1\rangle-\frac{i}{2}l_4|\xi a^2\rangle
\nonumber\\
\mathbbm T|a^1\chi\rangle&=&\frac12(-l_6-l_7+2c)|a^1\chi\rangle+\frac12 l_5|\chi a^1\rangle+\frac{i}{2}l_4|\chi a^2\rangle
\nonumber\\
\mathbbm T|a^2\chi\rangle&=&\frac12(l_6+l_7+2c)|a^2\chi\rangle-\frac12 l_5|\chi a^2\rangle+\frac{i}{2}l_4|\chi a^1\rangle
\nonumber\\
\mathbbm T|a^2\xi\rangle&=&\frac12(l_6+l_7+2c)|a^2\xi\rangle-\frac12 l_5|\xi a^2\rangle-\frac{i}{2}l_4|\xi a^1\rangle
\end{eqnarray}
\subsubsection*{Fermion-Fermion}
\begin{eqnarray}
\mathbbm T|\chi\chi\rangle=c|\chi\chi\rangle && \mathbbm T|\chi\xi\rangle=\frac12l_4|a^1a^1\rangle-\frac12l_4|a^2a^2\rangle-\frac{i}{2}l_5|a^1a^2\rangle-\frac{i}{2}l_5|a^2a^1\rangle
\nonumber\\
\mathbbm T|\xi\xi\rangle=c|\xi\xi\rangle &&
\mathbbm T|\xi\chi\rangle=\frac12l_4|a^1a^1\rangle-\frac12l_4|a^2a^2\rangle+\frac{i}{2}l_5|a^1a^2\rangle+\frac{i}{2}l_5|a^2a^1\rangle\,,
\end{eqnarray}
where
\bea
&&l_1=\frac{1}{2\hat{g}}\frac{(p_1+p_2)^2}{\omega_2 p_1-\omega_1 p_2}\,,\qquad l_2=\frac{1}{2\hat{g}}\frac{(p_1-p_2)^2}{\omega_2 p_1-\omega_1 p_2}\,,\qquad l_3=-\frac{1}{2\hat{g}}\frac{p_1^2-p_2^2}{\omega_2 p_1-\omega_1 p_2}\,, \\ \nn
&&
l_4=-\frac{1}{2\hat{g}}\frac{p_1 p_2}{p_1+p_2}\big(\sqrt{(\omega_1+p_1)(\omega_2+p_2)}-\sqrt{(\omega_1-p_1)(\omega_2-p_2)}\big),\\ \nn
&&
l_5=-\frac{1}{2\hat{g}}\frac{p_1 p_2}{p_1-p_2}\big(\sqrt{(\omega_1+p_1)(\omega_2+p_2)}+\sqrt{(\omega_1-p_1)(\omega_2-p_2)}\big),\\ \nn
&& l_6=\frac{1}{2\hat{g}}\frac{(p_1+p_2)p_2}{\omega_2 p_1-\omega_1 p_2}\,,\qquad l_7=-\frac{1}{2\hat{g}}\frac{(p_1-p_2)p_2}{\omega_2 p_1-\omega_1 p_2}\,.
\eea
When $c=0$, i.e. in the $a=\frac12$ uniform light-cone gauge used in this paper, this agrees exactly with the tree-level S-matrix computed in section \ref{sec:massive-massive}.

This is of course not expected to hold beyond tree-level. In fact from the exact S-matrix proposed in \cite{Borsato:2013hoa} we have for the ${a^1 a^2}\rightarrow {a^1 a^2}$ process at one loop
\begin{eqnarray}
\big(B_{LL}^2+C_{LR}^2\big)\vert_{1-loop}=-\frac{1}{2 h^2}\Big(\frac{p_1 p_1}{\omega_1 p_2-\omega_2 p_1}\Big)^2\Big[\frac{1}{2}\frac{\big(p_1^2-p_2^2\big)^2}{\big(p_1 p_2\big)^2}+\big(1-p_1p_2+\omega_1 \omega_2\big)\Big]\,.
\end{eqnarray}
and where the tree-level part coincides with $\ell_2^{(0)}$.   While the above resembles the worldsheet result there is an overall factor of 1/2 wrong in front of the first term in the bracket. It is furthermore straightforward to check that the amplitude for ${a^1 a^1}\rightarrow {a^1 a^1}$ given by $A_{LL}^2+A_{LR}^2$ also fails to reproduce the NFS worldsheet result.

\bibliographystyle{my-JHEP-4}
\bibliography{complete-library-processed,ads2ads3etc} %

\end{document}